\documentclass[prb,aps,floatfix]{revtex4}
\usepackage{graphicx,graphics,color,epsfig}
\usepackage{amsmath}
\usepackage{color}
\usepackage{amsfonts}
\usepackage{amssymb}
\usepackage{bm}

\begin{document}

\title{Majorana Fermions on Zigzag Edge of Monolayer Transition Metal
Dichalcogenides}
\author{Rui-Lin Chu$^{1}$}
\author{Gui-Bin Liu$^{2,3}$}
\author{Wang Yao$^{3}$}
\author{Xiaodong Xu$^{4,5}$}
\author{Di Xiao$^{6}$}
\author{Chuanwei Zhang$^{1}$}
\thanks{chuanwei.zhang@utdallas.edu}
\affiliation{$^{1}$ Department of Physics, the University of Texas at Dallas, Richardson,
TX 75080 USA \\
$^{2}$ School of Physics, Beijing Institute of Technology, Beijing 100081,
China \\
$^{3}$ Department of Physics and Center of Theoretical and Computational
Physics, The University of Hong Kong, Hong Kong, China \\
$^{4}$ Department of Physics, University of Washington, Seattle, Washington,
USA\\
$^{5}$ Department of Material Science and Engineering, University of
Washington, Seattle, Washington, USA \\
$^{6}$ Department of Physics, Carnegie Mellon University, Pittsburgh,
Pennsylvania 15213, USA }

\begin{abstract}
Majorana fermions, quantum particles with non-Abelian exchange statistics,
are not only of fundamental importance, but also building blocks for
fault-tolerant quantum computation. Although certain experimental
breakthroughs for observing Majorana fermions have been made recently, their
conclusive dection is still challenging due to the lack of proper material
properties of the underlined experimental systems. Here we propose a new
platform for Majorana fermions based on edge states of certain
non-topological two-dimensional semiconductors with strong spin-orbit
coupling, such as monolayer group-VI transition metal dichalcogenides (TMD).
Using first-principles calculations and tight-binding modeling, we show that
zigzag edges of monolayer TMD can host well isolated single edge band with
strong spin-orbit coupling energy. Combining with proximity induced \textit{s%
}-wave superconductivity and in-plane magnetic fields, the zigzag edge
supports robust topological Majorana bound states at the edge ends, although
the two-dimensional bulk itself is non-topological. Our findings points to a
controllable and integrable platform for searching and manipulating Majorana
fermions.
\end{abstract}

\maketitle

Majorana fermions\cite{Wilczek} are quantum particles that are their own
anti-particles, and satisfy non-Abelian exchange statistics. The latter is
the key for their potential use in fault-tolerant topological quantum
computation\cite{TQC}, which makes their experimental realization an
extremely important task from the long-term technological perspective. In
the past two decades, some exotic condensed matter/cold atom systems\cite%
{TQC,Fuliang-majo,CW,Sau-prl,Lutchyn-prl,Oreg-prl,Alicea-prb,Mao} have been
proposed to support Majorana fermions. The experimental breakthrough occurs
only recently\cite{Kouwenhoven,Deng-nano,Das-nat,Rokhinson}, using
heterostructures consisting of conventional s-wave superconductors and
semiconducting nanowires subjected to an external magnetic field, where
certain signatures of Majorana fermions were observed. However, there are a
few material complications inherent to semiconductor nanowires that may
prevent the experimental signature from being conclusive\cite%
{Liujie-prl,Altland-prl, Kells-prb,JHLee-prl,Pikulin-njp,Sarma-rapid}: (i)
The large diameters of the nanowires yield multiple occupied transversal sub
bands, resulting in complications for the superconductor proximity effect
and the chemical potential level\cite%
{Wimmer-10prl,Potter-10prl,Lutchyn-11prl}; (ii) The spin-orbit coupling in
these wires is rather weak, which renders the Majorana physics extremely
vulnerable to disorder, making it challenging to exclude alternative
interpretations of the experimental signature based on disorder effect\cite%
{Stanescu-11prb,Potter-11prb,Liujie-prl}; (iii) The random growth process of
nanowires also makes it hard to build a nanowire network\cite{Alicea-nat} to
detect the statistics of Majorana fermions.\newline

Amid the above difficulty, it is critically important to look for other 1D
conducting states to realize Majorana fermions. A natural and more
controllable way is to consider 1D edge states of a 2D material. In this
context, helical edge states of 2D quantum spin Hall insulators (QSHIs) have
been proposed to support Majorana fermions\cite{Fuliang-qshi-majo,du1,xlq}.
However, so far the QSHIs have only been realized in semiconductor
heterostructures and are subjected to stringent growth conditions.
Furthermore, the bulk itself of a QSHI is generally not a good insulator
because of the relatively small band gap (about 10 meV). It is therefore
natural to ask whether the edge states of non-topological 2D materials with
a large bulk band gap can support Majorana fermions. While in pursuit of
such platforms a few key material properties are of particular interest: (i)
The compounds must have heavy elements that can generate strong spin-orbit
coupling (SOC) necessary for robust 1D topological superconductors\cite%
{Potter-12prb,sau2}; and (ii) 2D atomically thin materials with
honeycomb-like lattice structures (\textit{i.e.}, similar as graphene),
which are more likely to support single band edge states.\newline

In this article we demonstrate this idea by showing that 1D zigzag edges of
a new class of 2D semiconductors, monolayer group-IV transition metal
dichalcogenides (TMDs), provide a promising new platform for studying 1D
topological superconductors with a single transversal band, strong SOC
energy, and controlled Majorana network generation. Using both
first-principles calculations and tight-binding modeling, we show that the
chalcogen-terminated zigzag edges of these 2D semiconductors support edge
bands with strong Rashba-type SOC and are well separated from the large bulk
bands ($\sim $1.5 to 1.8 eV). By utilizing a minimal realistic tight-binding
model, we numerically confirm the existence of zero-energy Majorana states
at the two ends of the edge in the presence of proximity induced \textit{s}%
-wave superconductivity, and their robustness against disorders. Our
findings point out a new pathway for searching for Majorana fermions using
edge states of widely existing 2D non-topological semiconductors.\newline

{\LARGE \textbf{Results}}\newline

\begin{figure}[tbph]
\centering
\includegraphics[width=0.95\textwidth]{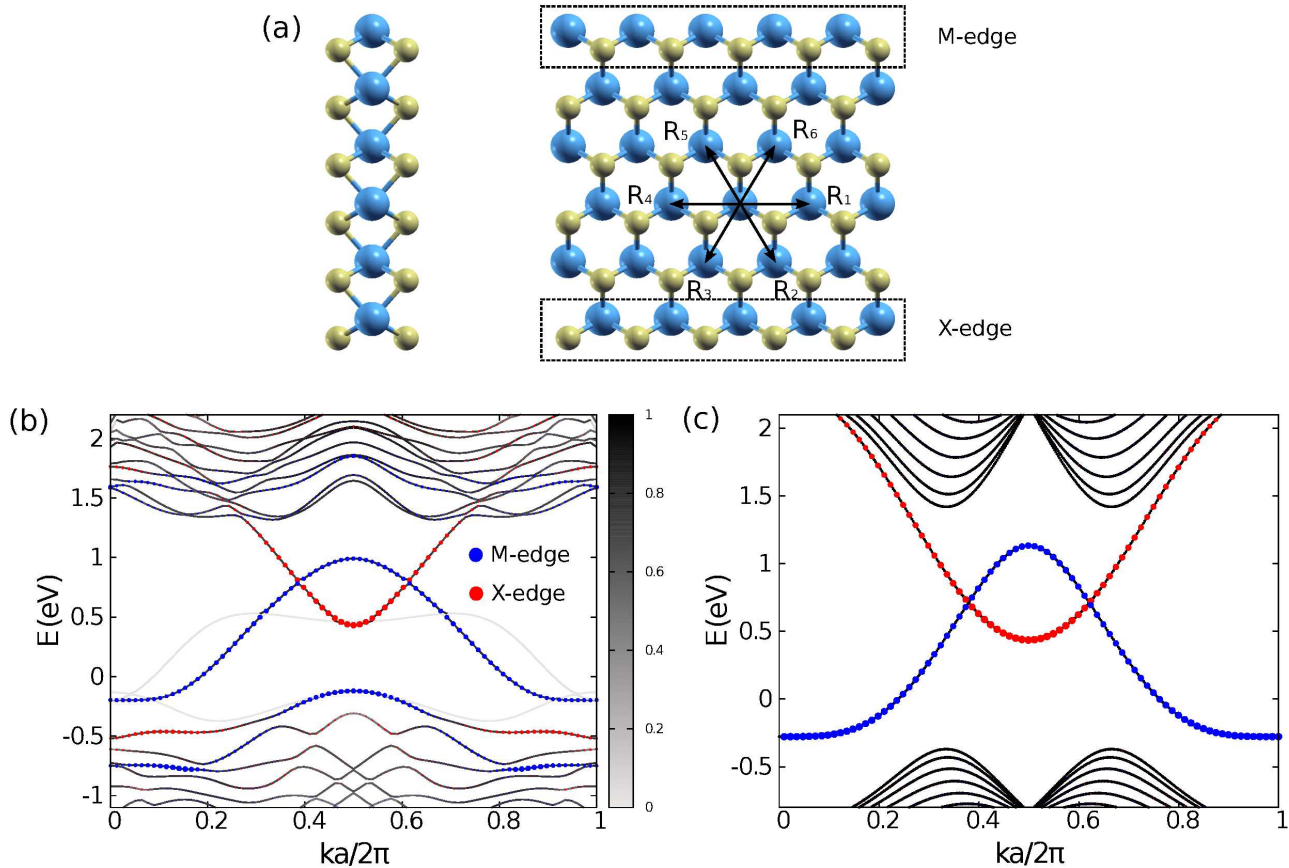}\newline
\caption{ Isolated single edge band in monolayer MX$_{2}$. (a) Side and top
view of a monolayer MX$_{2}$ zigzag ribbon. $R_{i}$ are the vectors
connecting the nearest M atoms. The ribbon is infinite in $R_{1}$ and $R_{4}$
directions. The lower and upper edges of the ribbon are referred to as
X-edge and M-edge, respectively. The ribbon's width is measured by the
number of zigzag chains $N_{c}$ in width. (b) DFT band structure of a WSe$%
_{2}$ zigzag ribbon with $N_{c}=8$, not including SOC. The grey scale bar
represents the total orbital ($d_{z^{2}}+d_{xy}+d_{x^{2}-y^{2}}$) weight of
the band. The color dots represent the orbital ($%
d_{z^{2}}+d_{xy}+d_{x^{2}-y^{2}}$) weight (with larger dot for larger
weight) from the M atoms on X-edge or M-edge. Three pairs of well localized
in-gap edge states can be identified, one (red dot) is on the X-edge and the
other two (blue dot and light gray) on the M-edge. The two inequivalent
valleys $K$ and $K^{\prime }$ are located at $ka/2\protect\pi =1/3$ and $2/3$%
, respectively. $a=3.31{\text{\normalfont\AA }}$ is the bulk's lattice
constant. Fermi surface is at $E=0$. See supplementary note for similar
plots for MoS$_{2}$, MoSe$_{2}$ and WS$_{2}$. (c) Same as (b) with
tight-binding model and $N_{c}=20$. }
\label{fig1}
\end{figure}

\textbf{Isolated single edge band.} We consider four different TMDs: MoS$%
_{2} $, MoSe$_{2}$, WS$_{2}$ and WSe$_{2}$, but will present our results in
the following mainly using WSe$_{2}$ as a representative because of its much
larger SOC energy. Monolayer TMDs are atomically thin 2D direct-bandgap
semiconductors with exotic coupled spin and valley physics~\cite{xiaodi-mo}
and excellent optical properties, as demonstrated in recent experiments\cite%
{Mak-10prl,AKis,cao-nat-com,Mak2,xiaodongcui,xu1,xu2}. Structurally
monolayer MX$_{2}$ is a tri-layer X-M-X sandwich. Within each layer, M and X
atoms form 2D hexagonal lattices. When viewed from top it shows a honeycomb
structure. The 2D bulk of monolayer MX$_{2}$ has a direct band gap of 1.5 $-$
1.8 eV located at the corners of its 2D hexagonal Brillouin zone called
valleys\cite{xiaodi-mo}. The bulk's edges can be classified as zigzag and
armchair types like in graphene. Due to the lacking of inversion symmetry in
the monolayer, the zigzag edges can be further classified as X-terminated
and M-terminated, which correspond to the lower and upper edges of the
ribbon shown in Fig. 1a. We refer to them as X-edge and M-edge respectively.
It is already known from STM measurements that the zigzag edges of
triangular shaped monolayer MoS$_{2}$ nanoflakes support multiple pairs of
1D metallic edge states\cite{Helveg}. The edges of these nanoflakes are
later identified as Mo-edge with passivated S atoms\cite{Bollinger}. For a
zigzag MX$_{2}$ ribbon shown in Fig. 1a, the edge states exist on both the
M-edge and X-edge.\newline

In addition to the density functional theory (DFT) calculations, insight
into the underlying physics can be obtained from a minimal tight-binding
model that is constructed by considering the lattice symmetry and the
corresponding crystal field splitting. It is known that the valence band
maximum and conduction band minimum of monolayer MX$_{2}$ consist mainly of
M atom's \textit{d} orbitals. Thus to describe the low energy band structure
of the monolayer's bulk it is sufficient to consider the \textit{d} orbitals
from the M atoms\cite{xiaodi-mo}. The trigonal prismatic coordination of the
M atom splits its \textit{d} orbitals into three groups: $A_{1}^{\prime
}(d_{z^{2}})$, $E^{\prime }(d_{xy}$, $d_{x^{2}-y^{2}}$) and $E^{\prime
\prime }(d_{xz}$, $d_{yz})$. The monolayer's mirror symmetry in the $\hat{z}$
direction permits hybridization only between the $A_{1}^{\prime }$ and $%
E^{\prime }$ groups. This allows us to consider three orbitals of $d_{z^{2}}
$, $d_{xy}$, and $d_{x^{2}-y^{2}}$ for a minimal tight-binding model. We
refer the readers to Ref\cite{Liuguibin} for detailed descriptions of this
tight-binding model including the symmetry analysis and material specific
parameters fitted from first-principle calculations. The tight-binding model
is able to capture the essential physics of the monolayer, including the
direct band gaps at the $K$ and $K^{\prime }$ valleys, the degeneracy of the
band edges and the valley contrast spin splitting of the valence band due to
SOC, etc\cite{xiaodi-mo,Liuguibin}.\newline

In simple languages this tight-binding model only considers M atoms' on-site
energies and electron hopping along the six vectors connecting the nearest M
atoms (marked as $R_{1}\sim R_{6}$ in Fig. 1a). Without considering SOC and
the spin degree of freedom, the tight-binding Hamiltonian can be written as
a $3\times 3$ matrix
\begin{equation*}
\mathcal{H}(\mathbf{k})=\left(
\begin{array}{ccc}
H_{11}^{11} & H_{11}^{12} & H_{12}^{12} \\
~ & H_{11}^{22} & H_{12}^{22} \\
c.c. & ~ & H_{22}^{22}%
\end{array}%
\right) ,
\end{equation*}%
in the $\mathbf{k}$-space. Here the basis is taken as $\{|\phi
_{1}^{1}\rangle =d_{z^{2}},|\phi _{1}^{2}\rangle =d_{xy},|\phi
_{2}^{2}\rangle =d_{x^{2}-y^{2}}\}$ and $H_{ij}^{\alpha \beta }$ represents
the matrix element between $|\phi _{i}^{\alpha }\rangle $ and $|\phi
_{j}^{\beta }\rangle $, and is obtained from the Fourier transformation of
the real space tunneling matrix between neighboring sites.\newline

\begin{figure}[tbph]
\centering
\includegraphics[width=0.99\textwidth]{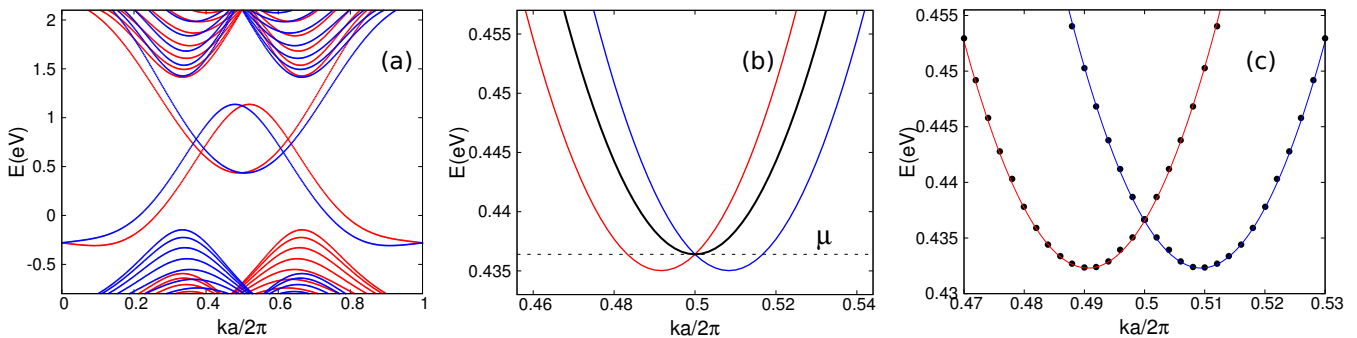}
\caption{X-Edge band minimum. (a) Tigh-binding band structure for zigzag
nanoribbon with SOC turned on. The color marks spin direction with red=spin
up, blue=spin down. $N_c=20$, $\protect\lambda =230meV$. (b) Zoom in of the
X-edge band minimum of (a) at $ka=\protect\pi $ . The black line is without
SOC. The dash line marks the chemical potential used in the BdG calculation
in Fig.3. (c) Same as (b) for Fig.1b with SOC turned on, where the dots are
DFT results and the lines represents the best fit from Eq.(\protect\ref%
{effectH}). See supplementary note for similar plots for MoS$_{2}$, MoSe$%
_{2} $ and WS$_{2}$.}
\end{figure}

\textbf{Rashba-type SOC in edge band.} We proceed to include SOC and
demonstrate how the Rashba-type SOC in the edge band is generated. The $%
\mathbf{L\cdot S}$ type SOC in MX$_{2}$ originates from the \textit{d}
orbitals of the heavy M atoms (Mo or W)\cite{zhuzhiyong,xiaodi-mo}. In
monolayer;s Bulk, the spi-oribt term can be described as
\begin{equation}
\mathcal{H}_{SO}=\frac{\lambda }{2}(S_{x}\otimes L_{x}+S_{y}\otimes
L_{y}+S_{z}\otimes L_{z}),
\end{equation}%
where $S_{i}$ and $L_{i}$ represent the spin and orbital angular momentum
operator respectively. It turns out in the basis of $\{|\phi _{1}^{1}\rangle
,|\phi _{1}^{2}\rangle ,|\phi _{2}^{2}\rangle \}$ the $L_{x}$ and $L_{y}$
are both $0_{3\times 3}$ matrices, which enables us to write the total
Hamiltonian in a spin-decoupled form
\begin{equation*}
\mathcal{H}^{\prime }(\mathbf{k})=I_{2}\otimes \mathcal{H}(\mathbf{k})+%
\mathcal{H}_{SO}=\left(
\begin{array}{cc}
\mathcal{H}(\mathbf{k})+\frac{\lambda }{4}L_{z} & 0 \\
0 & \mathcal{H}(\mathbf{k})-\frac{\lambda }{4}L_{z} \\
&
\end{array}%
\right) ,~~~L_{z}=\left(
\begin{array}{ccc}
0 & 0 & 0 \\
0 & 0 & 2i \\
0 & -2i & 0%
\end{array}%
\right) .
\end{equation*}%
Here the upper and lower sub-blocks represent the spin up and down
respectively. It is noted that $\mathcal{H}(\mathbf{k})$ is a
time-reversal-invariant Bloch Hamiltonian meaning $\Theta \mathcal{H}(%
\mathbf{k})\Theta ^{-1}=\mathcal{H}(-\mathbf{k})$, where $\Theta $ is the
time-reversal operator. The total Hamiltonian $\mathcal{H}^{\prime }(\mathbf{%
k})$ also respects the time-reversal symmetry(TRS) in this sense. However,
when view each spin block individually, the TRS is explicitly broken by the
SOC term $\pm \frac{\lambda }{4}L_{z}$.\newline

In Figs. 1b and 1c and Fig. 2 we present both the DFT and tight-binding band
structure of the WSe$_{2}$ zigzag ribbon, where the edge states localized on
different edges are marked correspondingly. The electrons from these edge
bands dominantly resides on the M atoms of the zigzag edges. Comparing with
the DFT band structure, we see the tight-binding model can successfully
captures the parabolic edge bands on both the X-edge and M-edge. It's worth
to mention that the effective SOC we find in the parabolic M-edge band is
generally larger than in the X-edge band. Nevertheless the X-edge is
preferable for two main reasons. First, X-edge hosts a single edge band
while the M-edge hosts multiple edge bands; Second, the X-edge is
structurally very stable while the M-edge can be dramatically affected by
edge passivations\cite{Seifert, PanHui, Wang-jacs,Chen-jacs}. Here after, we
will focus on the X-edge band.\newline

In Fig.2b we compare the X-edge band before and after turning on the SOC.
Apparently the $\mathcal{H}_{SO}$ can be viewed as a TRS breaking
perturbation term, whose effect is slightly shifting the spin up branch to
the left and spin down branch to the right. The whole band structure
nevertheless remains symmetric about $ka=\pi $ because of the TRS.
Accordingly the low energy effective 1D Hamiltonian for the X-edge band can
be written as
\begin{equation}
\mathcal{H}_{eff}(k^{\prime })=\frac{1}{2m^{\ast }}k^{\prime 2}+\alpha
_{R}k^{\prime }\sigma ^{z}+C,  \label{effectH}
\end{equation}%
where $\sigma ^{z}$ is the $z$ component of the Pauli matrix, $k^{\prime
}=k-\pi /a$, and $C$ is a constant. Up to a unitary transformation, this
Hamiltonian is equivalent to that for the semiconductor nanowires with
Rashba-type SOC\cite%
{Lutchyn-prl,Alicea-nat,Wimmer-10prl,Potter-10prl,Lutchyn-11prl,Kouwenhoven,Deng-nano,Das-nat}%
. Here $\alpha _{R}$ is the effective Rashba velocity. In Table \ref{table1}
the effective mass and $\alpha _{R}$ fitted from our first-principle
calculations for MoS$_{2}$, MoSe$_{2}$, WS$_{2}$ and WSe$_{2}$ are listed.
These parameters are orders of magnitude larger than their semiconductor
nanowire counterparts, especially for WS$_{2}$ and WSe$_{2}$.\newline

\begin{table}[h]
\caption{Effective mass and Rashba velocity for TMD's S(Se)-edge band and
semiconductor nanowires. SOC energy is defined as $E_{SO}=\frac{1}{2}m^{\ast
}\protect\alpha _{R}^{2}$. }
\label{table1}%
\begin{tabular}{|c|c|c|c|c|c|c|}
\hline
~ & MoS$_2$ & MoSe$_2$ & WS$_2$ & WSe$_2$ & InAs\cite{Alicea-prb} & InSb\cite%
{Kouwenhoven} \\ \hline
$m^*(m_e)$ & 0.28 & 0.24 & 0.33 & 0.31 & 0.04 & 0.015 \\ \hline
$\alpha_R(eV{\text{\normalfont\AA }})$ & 0.12 & 0.11 & 0.33 & 0.46 & 0.06 &
0.2 \\ \hline
$E_{SO}(meV)$ & 0.26 & 0.2 & 2.3 & 4.3 & 0.01 & 0.05 \\ \hline
\end{tabular}%
\end{table}

\begin{figure}[tbph]
\centering
\includegraphics[width=0.9\textwidth]{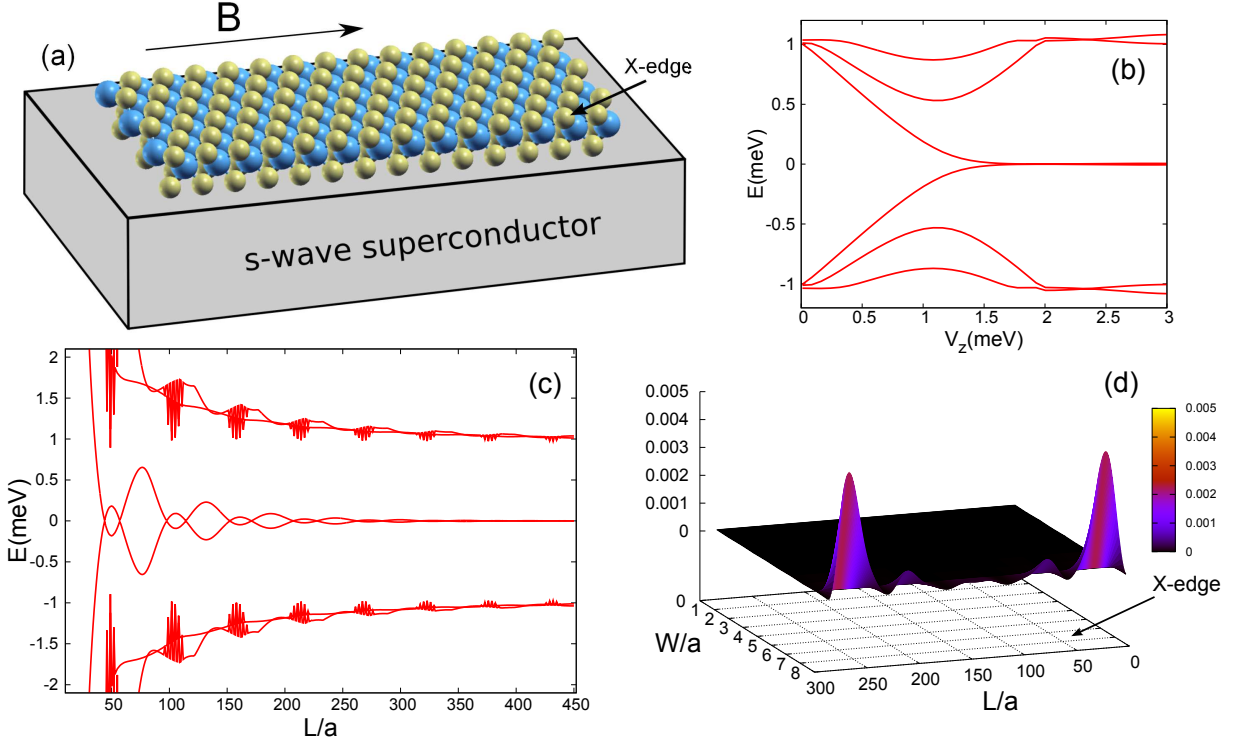}\newline
\caption{ Majorana zero-energy mode. (a) A monolayer MX$_{2}$ zigzag ribbon
deposited on top of an s-wave superconductor. A top gate can be applied to
tune the chemical potential. An external magnetic field is applied parallel
to the zigzag edge in order to make the system a topological superconductor.
(b) The emergence of the zero-energy mode. $L/a=400$; (c) Evolution of low
energy spectrum with ribbon's length $L$. $V_{z}=2.0~meV$. In (b) and (c)
only 6 modes closest to zero-energy are plotted. (d) Real-space distribution
of the zero-energy mode over the ribbon for $L/a=300$. The 3-D view angle is
set to be the same as that in (a). Other parameters are: $N_{c}=10$, $\Delta
=1.0~meV$, $\protect\mu =0.4364eV$. }
\end{figure}

\textbf{Existence of Majorana end states.} To create a 1D topological
superconductor, we introduce superconducting pairing through proximity
effects by depositing the MX$_{2}$ monolayer on top of a conventional
\textit{s}-wave superconductor (Nb, NbSe$_{2}$, etc.), as illustrated in
Fig. 3a. The MX$_{2}$ monolayers have X-M-X layer thickness $\approx $ 3 ${%
\text{\normalfont\AA }}$, which is well within the superconducting coherence
length of a typical \textit{s}-wave superconductor. The selected monolayer
can either be (1) a zigzag nanoribbon or (2) a large monolayer sample with
an identified X-edge. A top gate can then be applied locally to tune the
chemical potential. We note that in the first case there are coexisting edge
states on the M-edge as shown in the ribbon's band structures (Figs. 1b and
1c). These M-edge states are well localized on the M-edge's M atoms. To
realize a topological superconducting state, the chemical potential $\mu $
need be tuned to be around the X-edge band bottom near $k=\pi /a$ (see Fig.
2b). The corresponding M-edge states at this chemical potential occur at
momenta far away from $k=\pi /a$. In that region, even number of Mo-edge
bands are cut at the Fermi surface, which do not affect the topological
properties of the system\cite{Lutchyn-prl}. Therefore the M-edge states do
not interfere with the topological superconducting state on the X-edge,
which is also confirmed in our following numerical simulations. Such
coexistence of edge bands is completely eliminated for the second case,
where the X-edge can be well isolated. For this case the gate is only
required to cover the selected segment of the edge since the bulk maintains
a large band gap.\newline

To demonstrate the functionality of the proposed setup, we carry out a
numerical simulation in 2D with the tight-binding model. We adopt the ribbon
structure for this purpose as illustrated in Fig. 3a. Because of their
excellent structural stability, TMD zigzag nanoribbons can be synthesized
with uniform width and smooth edges without defects \cite%
{Wang-jacs,Wang-JMC,ZhengLiu}. To drive the system into a topological
superconducting state, an in-plane magnetic field is applied to create a
Zeeman splitting gap at the band crossing point.\newline

The Zeeman term induced by the magnetic field reads
\begin{equation}
\mathcal{H}_{Z}=V_{z}\sum_{i,l\alpha \beta }c_{i,l\alpha }^{\dagger }\sigma
_{\alpha ,\beta }^{x}c_{i,l\beta },
\end{equation}%
where $c_{i,l\alpha }^{\dagger }$ is the creation operator for electron on
site $i$ with orbital index $l$ $(1\sim 3)$ and spin index $\alpha $ and $%
\beta $. We have assumed the magnetic field is in the $x$ direction, but
nevertheless any in-plane magnetic field would work equivalently. The
proximity effect induced superconducting paring term writes
\begin{equation}
\mathcal{H}_{SC}=\sum_{i,l}(\Delta c_{i,l\uparrow }^{\dagger
}c_{i,l\downarrow }^{\dagger }+h.c.),
\end{equation}%
where for simplicity we assume a uniform intra-orbit pairing strength.
Denote the lattice version of the ribbon's Hamiltonian in Eq. \ref{tot-H} as
$\mathcal{H}_{0}$, we then solve the corresponding BdG equation for the full
Hamiltonian (see Methods)
\begin{equation}
\mathcal{H}=\mathcal{H}_{0}-\mu +\mathcal{H}_{Z}+\mathcal{H}_{SC}
\end{equation}%
to get the low energy spectrum. The emergence of the zero-energy mode with
increasing Zeeman splitting is shown in Fig. 3b. In Fig. 3c we show the
evolution of the 6 lowest energy modes with the ribbon's length. As a
signature of Majorana fermions\cite{Tudor-13prb}, the zero-energy modes show
an oscillating energy splitting with an exponentially decaying envelope.
When the ribbon is sufficiently long the zero-energy mode as well as the
excitation gap become well defined. We notice that the fast alternating
modes in the excitation gap of a short ribbon are contributions from the
coexisting M-edge bands, which confirms our prediction that they do not
affect the topological superconductor on the X-edge. It is well established
that in a 1D topological superconductor the Majorana fermions appear as end
states in real space\cite{Sau-prl,Lutchyn-prl}. We confirm this by plotting
the particle component of the zero-energy mode wave function in Fig.3d,
where the ribbon's size is $50~nm$ in length and $2.5~nm$ in width.\newline

\begin{figure}[tbph]
\centering
\includegraphics[width=0.8\textwidth]{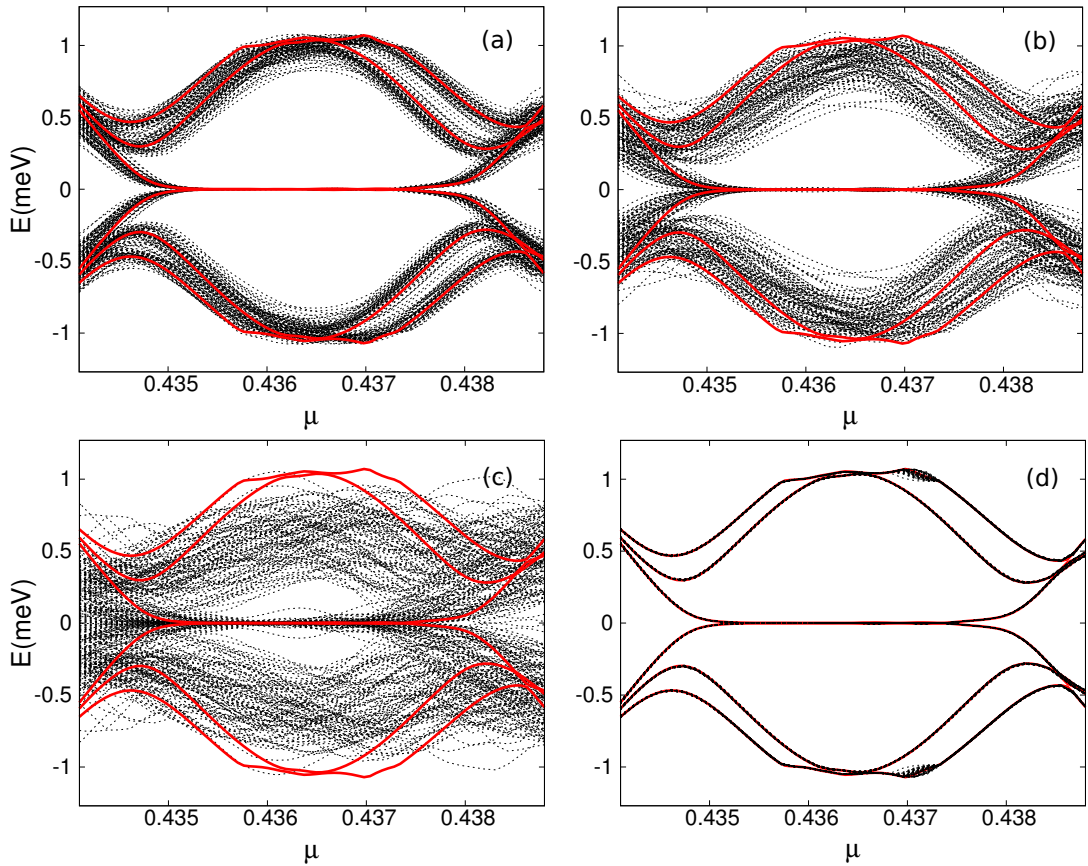}\newline
\caption{ Effect of edge and bulk disorders. In each panel 50 random
disorder configurations are collected and the 6 lowest energy modes for each
disorder configuration are plotted. Red and black curve represent clean and
disordered case respectively. (a-c) Disorders are put on both the ribbon's
bulk and edge. (a) $W=5meV$; (b) $W=10meV$; (c) $W=20meV$; (d) Disorders are
put only on the ribbon's bulk but not on the edge, $W=200meV$. $\Delta=1mev$%
, $V_z=2meV$, $L/a=400$.}
\end{figure}

\textbf{Effect of disorder.} It is important that Majorana fermions can
sustain certain amount of disorders since in realistic experimental
conditions disorders are unavoidable. To explore the disorder effect in this
system, we add random on-site potential
\begin{equation}
\mathcal{H}_{dis}=\sum_{i,l\alpha }\varepsilon _{i}c_{i,l\alpha }^{\dagger
}c_{i,l\alpha },
\end{equation}%
to the tight-binding model, where $\varepsilon _{i}$ are normally
distributed in the range [-$W/2$,$W/2$]. We have simulated two kinds of
disorders: (1) disorder covers both bulk and edge; (2) disorder only covers
the bulk but not the edge. As shown in Fig.4a-c, the zero-energy modes as
well as the excitation gap are robust against edge disorders up to $W\sim
10\Delta $ ($W/\Delta $ can be much larger for a smaller $\Delta $). For
even stronger edge disorders the excitation gap starts to diminish and the
zero-energy modes gain splitting. An important advantage of this proposed
system is that the topological superconductor resides only on the edge, as a
result it gains strong immunity from bulk disorders. This is demonstrated in
Fig.4d where the zero-energy modes and excitation gap remain totally intact
despite of the strong disorder in the bulk.\newline

\textbf{Material considerations.} Although both the zigzag and armchair
edges of monolayer MX$_{2}$ support edge states, the zigzag edges are
generally more stable than the armchair edges. Consequently during
nanoribbon synthesis the zigzag nanoribbons dominates\cite%
{Wang-jacs,Wang-JMC,ZhengLiu}. In particular, the X-edge shows maximal
stability among all the edge configurations of monolayer MX$_{2}$\cite%
{Wang-jacs,PanHui,Chen-jacs}. In the experimentally grown WS$_{\mathbf{2}}$\
nanoribbon, the S-edge was found to be perfect without defects \cite%
{ZhengLiu}. Furthermore, edge states have been observed in MoS$_{2}$
nanoflakes using STM\cite{Helveg,Bollinger}.\newline

In the DFT band structures of MX$_{2}$ zigzag ribbon (Fig.1b and
supplementary note), three pairs of in-gap edge states exist connecting the
two inequivalent valleys. All of them consists dominantly of the \textit{d}
orbitals from the M atoms of the ribbon's outermost zigzag chain. Notably
two of these edge states have parabolic dispersions at their band
minimum/maximum near $ka=\pi $, which suits our purpose to find an effective
Hamiltonian like Eq.\ref{effectH}. The two parabolic bands are also fully
captured in the tight-binding model. The edge state phenomenon is similar in
all MX$_{2}$ zigzag ribbons we have calculated. After including SOC, the
Rashba-type SOC in the edge band is evident when zooming in the band bottom
at $ka=\pi $ (Fig.2c and supplementary note)). We have also calculated
ribbons with different width. We find these well localized edge states start
to exist in very narrow ribbons ($N_{c}=4$).\newline

The Fermi surface in DFT calculations of suspended MX$_{2}$ is typically in
the band gap and close to the valence band top. However, both n-type\cite%
{AKis,Xu-acsnano,Zande-natm} and p-type\cite{ptype1,ptype2} conductivities
have been reported in transport measurements, suggesting a wide-range
tunability for the chemical potentials\cite{Dolui}. The material synthesis
and device fabrication of monolayer MX$_{2}$ are rapidly developing because
of their potential applications in the next generation of electronics. The
nanoribbon samples for a possible experimental realization for our proposed
setup have become readily available\cite{Wang-jacs,Wang-JMC,ZhengLiu}. The
monolayer's 2D nature and similarity with graphene also makes many well
developed 2D device engineering and fabrication techniques directly
applicable.\newline

{\LARGE \textbf{Discussion}}\newline

The advantage of the proposed platform can be summarized in such a few
aspects: (1) The single edge band is well isolated in the middle of a
relatively large bulk band gap, which would lead to a minimal background
signal in the zero-bias peak measurements for detecting Majorana fermions.
The well known multiband problem in the semiconductor nanowires is
successfully avoided\cite{Potter-10prl}. More importantly, the platform is a
true 1D system localizing on a single atomic chain resulting in strong
immunity from bulk's disorders as we have demonstrated; (2) The system is
atomically thick, which would lead to robust and uniform superconducting
proximity effect when placed on top of the s-wave superconductor. It would
also result in efficient gate tunability, which has already been
demonstrated in transport measurements with the monolayer\cite{AKis,YZhang};
(3) The large effective mass and large effective Rashba-SOC in this platform
is unparalleled to the conventional semiconductor nanowires. As a result,
the proximity induced superconducting pairing and the associated Majorana
fermions are robust against disorders\cite%
{Stanescu-11prb,Potter-11prb,Potter-12prb,sau2}; (4) The 1D edge states can
be controllably generated from 2D materials, making it possible to construct
a Majorana network\cite{Alicea-nat} for studying the non-Abelian statistics
of Majorana fermions and implementing topological quantum computation.\newline

{\LARGE \textbf{Methods}}\newline

\textbf{Real-space BdG equation.} Here we elaborate how the BdG calculation
for Fig. 3 is implemented on the zigzag ribbon's lattice. Denote the
spin-independent $3\times 3$ hopping matrix from the nearest M atoms at site
$i$ to $j$ as $T_{ij}$ and the on-site potential matrices as $H_{on}=$ diag$%
(\epsilon _{1}~\epsilon _{2}~\epsilon _{2})$\cite{Liuguibin}. Then the
real-space BdG equation can be written as
\begin{equation*}
\sum_{j}\left(
\begin{array}{cccc}
H_{\uparrow \uparrow } & 0 & 0 & \Delta ^{\prime } \\
0 & H_{\downarrow \downarrow } & -\Delta ^{\prime } & 0 \\
0 & -\Delta ^{\prime \ast } & -H_{\uparrow \uparrow }^{\ast } & 0 \\
\Delta ^{\prime \ast } & 0 & 0 & -H_{\downarrow \downarrow }^{\ast }%
\end{array}%
\right) _{ij}\left(
\begin{array}{c}
u_{nj}^{\uparrow } \\
u_{nj}^{\downarrow } \\
v_{nj}^{\uparrow } \\
v_{nj}^{\downarrow }%
\end{array}%
\right) =\varepsilon _{n}\left(
\begin{array}{c}
u_{ni}^{\uparrow } \\
u_{ni}^{\downarrow } \\
v_{ni}^{\uparrow } \\
v_{ni}^{\downarrow }%
\end{array}%
\right) ,
\end{equation*}%
where $\sigma \equiv \{\uparrow ,\downarrow \}$, $H_{\sigma \sigma
,ij}=T_{ij}$ when $i\neq j$ and $H_{\sigma \sigma ,ii}=H_{on}\pm \frac{%
\lambda }{4}L_{z}$, $\Delta ^{\prime }=\delta _{ij}\cdot I_{3}\otimes \Delta
$. $u_{ni}^{\sigma }$ and $v_{ni}^{\sigma }$ each being a $3\times 1$ vector
are the components of the $n-$th quasiparticle wave function at site $i$, $%
\varepsilon _{n}$ is the corresponding energy eigenvalue. The low energy
spectrum and wave function is then obtained by using the sparse matrix
eigensolver in MATLAB.\newline

\textbf{DFT calculation}. Our first-principle DFT calculations are performed
using the all-electron full-potential linearized augmented-planewave
(FP-LAPW) method\cite{ELK}. The SOC is included in terms of the
second-variational method with scalar-relativistic orbitals as a basis.
Construction of our tight-binding model is described in details in Ref\cite%
{Liuguibin}.\newline

\textbf{Acknowledgements}

R.C., and C.Z. are supported by ARO (W911NF-12-1-0334), AFOSR
(FA9550-13-1-0045), and NSF-PHY (1104546). G.L acknowledges support by the
Basic Research Funds of Beijing Institute of Technology with Grant No.
20121842003 and the National Basic Research Program of China 973 Program
with Grant No. 2013CB934500. W.Y. acknowledges
support by HKSAR Research Grant Council (HKU705513P), and Croucher
Foundation under the Croucher Innovation Award. X.X is supported by US DoE, BES, Division of Materials Sciences and Engineering (DE-SC0008145). D.X. acknowledges support by the U.S.
Department of Energy, Office of Basic Energy Sciences, Materials Sciences
and Engineering Division, and AFOSR (FA9550-12-1-0479).\newline

\textbf{Author contributions}

All authors made critical contributions to the work.\newline

\textbf{Competing financial interests}

The authors declare no competing financial interests.\pagebreak

\centering{\LARGE \textbf{Supplementary Materials}}

\subsection{Isolated S(Se)-edge band in MoSe$_{2}$, WS$_{2}$ and WSe$_{2}$
from DFT calculations}

\begin{figure}[tbph]
\centering
\includegraphics[width=0.99\textwidth]{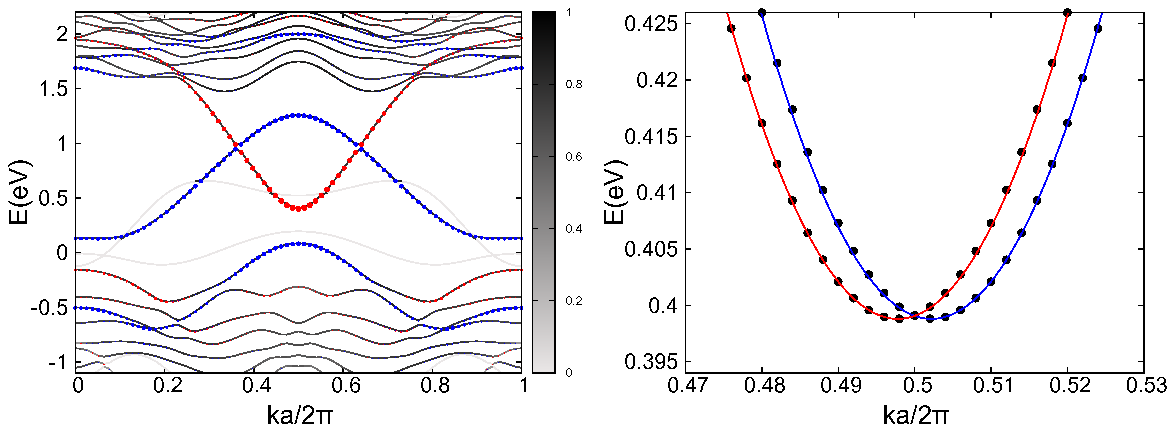}\newline
\caption{Left: DFT MoS$_{2}$ zigzag ribbon band structure, without SOC;
Right: Zoom in of the Se-edge band bottom, with SOC. }
\end{figure}
\begin{figure}[tbph]
\centering
\includegraphics[width=0.99\textwidth]{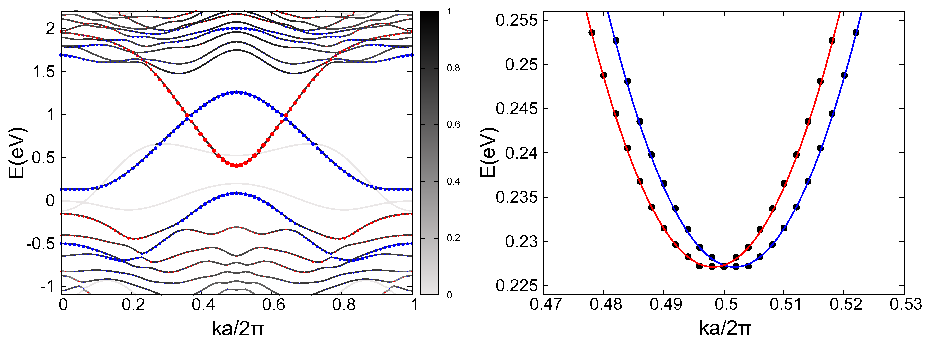}\newline
\caption{ Left: DFT MoSe$_{2}$ zigzag ribbon band structure, without SOC;
Right: Zoom in of the Se-edge band bottom, with SOC. }
\end{figure}
\begin{figure}[tbph]
\centering
\includegraphics[width=0.99\textwidth]{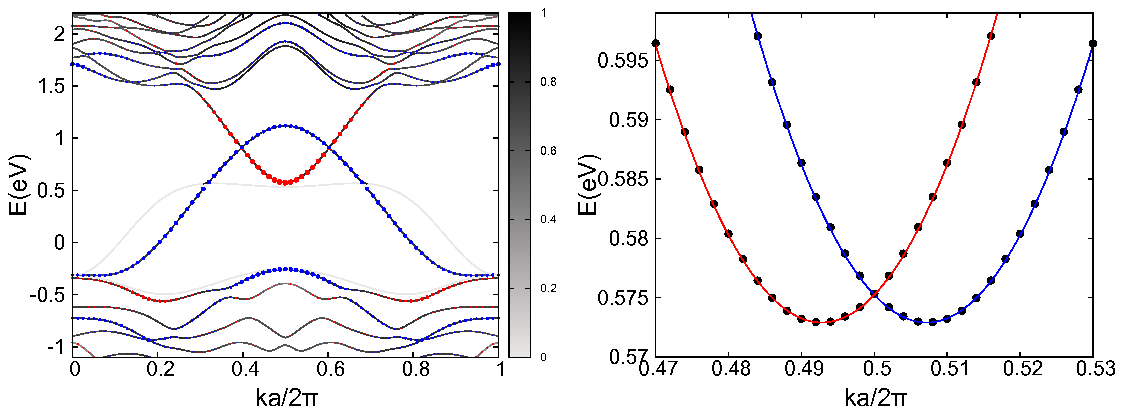}\newline
\caption{ Left: DFT WS$_{2}$ zigzag ribbon band structure, without SOC;
Right: Zoom in of the S-edge band bottom, with SOC. }
\end{figure}


\begin{thebibliography}{99}
\bibitem{Wilczek} Wilczek, F. Majorana returns. \textit{Nature Phys}.
\textbf{5,} 614-618 (2009).

\bibitem{TQC} Nayak, C., Simon, S. H., Stern, A., Freedman, M. \& Das Sarma,
S. Non-abelian anyons and topological quantum computation. \textit{Rev. Mod.
Phys.} \textbf{80,} 1083 (2008).

\bibitem{Fuliang-majo} Fu, L. \& Kane, C. L., Superconducting proximity
effect and Majorana fermions at the surface of a topological insulator.
\textit{Phys. Rev. Lett.} \textbf{100,} 096407 (2008).

\bibitem{CW} Zhang, C., Tewari, S., Lutchyn, R. M. \& Das Sarma, S. $%
p_{x}+ip_{y}$ superfluid from $s$-save interactions of fermionic cold atoms.
\textit{Phys. Rev. Lett.} \textbf{101,} 160401 (2008).

\bibitem{Sau-prl} Sau, J. D., Lutchyn, R. M., Tewari, S. \& Das Sarma, S.
Generic new platform for topological quantum computation using semiconductor
heterostructures. \textit{Phys. Rev. Lett.} \textbf{104,} 040502 (2010).

\bibitem{Lutchyn-prl} Lutchyn, R. M., Sau, J. D. \& Das Sarma, S. Majorana
fermions and a topological phase transition in semiconductor-superconductor
heterostructures. \textit{Phys. Rev. Lett.} \textbf{105,} 077001 (2010).

\bibitem{Oreg-prl} Oreg, Y., Refael, G. \& Oppen, F. V. Helical liquids and
Majorana bound states in quantum wires. \textit{Phys. Rev. Lett.} \textbf{%
105,} 177002 (2010).

\bibitem{Alicea-prb} Alicea, J. Majorana fermions in a tunable semiconductor
device. \textit{Phys. Rev. B} 81, 125318 (2010).

\bibitem{Mao} Mao, L., Gong, M., Dumitrescu, E., Tewari, S. \& Zhang, C.
Hole-doped semiconductor nanowire on top of an $s$-wave superconductor: a
new and experimentally accessible system for Majorana fermions. \textit{%
Phys. Rev. Lett.} \textbf{108,} 177001 (2012).

\bibitem{Kouwenhoven} Mourik, V. \textit{et al.} Signatures of Majorana
fermions in hybrid superconductor-semiconductor nanowire devices. \textit{%
Science} \textbf{336,} 1003-1007 (2012).

\bibitem{Deng-nano} Deng, M. T. \textit{et al.} Anomalous zero-bias
conductance peak in a Nb-InSb nanowire-Nb hybrid device. \textit{Nano Lett.}
\textbf{12,} 6414-6419 (2012).

\bibitem{Das-nat} Das, A. \textit{et al.} Zero-bias peaks and splitting in
an Al-InAs nanowire topological superconductor as a signature of Majorana
fermions. \textit{Nature Phys.} \textbf{8,} 887-895 (2012).

\bibitem{Rokhinson} Rokhinson, L. P., Liu, X. \& Furdyna, J. K. The
fractional a.c. Josephson effect in a semiconductor-superconductor nanowire
as a signature of Majorana particles. \textit{Nature Phys.} \textbf{8,}
795-799 (2012).

\bibitem{Liujie-prl} Liu, J., Potter, A. C., Law, K. T., and Lee, P. A.,
Zero-Bias Peaks in the Tunneling Conductance of Spin-Orbit-Coupled
Superconducting Wires with and without Majorana End-States, \textit{Phys.
Rev. Lett.} \textbf{109}, 267002 (2012).

\bibitem{Altland-prl} Bagrets, D., \& Altland, A. Class D Spectral Peak in
Majorana Quantum Wires, \textit{Phys. Rev. Lett.} \textbf{109}, 227005
(2012).

\bibitem{Kells-prb} Pientka, F., Kells, G., Romito, A., Brouwer, P. W., and
von Oppen, F., Enhanced Zero-Bias Majorana Peak in the Differential
Tunneling Conductance of Disordered Multisubband Quantum-Wire/Superconductor
Junctions, \textit{Phys. Rev. Lett.} \textbf{109}, 227006 (2012).

\bibitem{JHLee-prl} Lee, E. J. H. \textit{et al.} Zero-Bias Anomaly in a
Nanowire Quantum Dot Coupled to Superconductors. \textit{Phys. Rev. Lett.}
\textbf{109}, 186802 (2012).

\bibitem{Pikulin-njp} Pikulin, D. I., Dahlhaus, J. P., Wimmer, M.,
Schomerus, H., and Beenakker, C.W. J., A zero-voltage conductance peak from
weak antilocalization in a Majorana nanowire, \textit{New J. Phys.} \textbf{%
14} 125011 (2012).

\bibitem{Sarma-rapid} Das Sarma, S., Sau, J. D., \& Stanescu, T. D.,
Splitting of the zero-bias conductance peak as smoking gun evidence for the
existence of the Majorana mode in a superconductor-semiconductor nanowire,
\textit{Phys. Rev. B} \textbf{86}, 220506 (2012).

\bibitem{Wimmer-10prl} Wimmer, M. \textit{et al.} Majorana Bound States
without Vortices in Topological Superconductors with Electrostatic Defects,
\textit{Phys. Rev. Lett.} \textbf{105}, 046803 (2010).

\bibitem{Potter-10prl} Potter, A. C. \& Lee, P. A. Multichannel
generalization of Kitaev's Majorana end states and a practical route to
realize them in thin films. \textit{Phys. Rev. Lett.} \textbf{105,} 227003
(2010).

\bibitem{Lutchyn-11prl} Lutchyn, R. M., Stanescu, T.D., and Das Sarma, S.
Search for Majorana Fermions in Multiband Semiconducting Nanowires. \textit{%
Phys. Rev. Lett.} \textbf{106}, 127001 (2011).

\bibitem{Stanescu-11prb} Stanescu, T. D., Lutchyn, R. M., and Das Sarma, S.,
Majorana fermions in semiconductor nanowires, \textit{Phys. Rev. B} \textbf{%
84}, 144522 (2011).

\bibitem{Potter-11prb} Potter A. C. \& Lee, P. A. Engineering a p+ip
superconductor: Comparison of topological insulator and Rashba
spin-orbit-coupled materials. \textit{Phys. Rev. B} \textbf{83}, 184520
(2011).

\bibitem{Alicea-nat} Alicea, J., Oreg, Y., Refael, G., Oppen, F. V. \&
Fisher, M. P. A. Non-Abelian statistics and topological quantum information
processing in 1D wire networks. \textit{Nature Phys.} \textbf{7,} 412-417
(2011).

\bibitem{Fuliang-qshi-majo}
Fu F. \& Kane, C. L., Josephson Current and Noise at a
Superconductor-Quantum Spin Hall Insulator-Superconductor Junction, \textit{%
Phys. Rev. B }\textbf{79}, 161408(R) (2009).

\bibitem{du1} Knez, I., Du, R.-R. \& Sullivan, G., Andreev Reflection of
Helical Edge Modes in InAs/GaSb Quantum Spin Hall Insulator, \textit{Phys.
Rev. Lett.} \textbf{109}, 186603 (2012).

\bibitem{xlq} Qi, X.-L. \& Zhang, S.-C., Topological insulators and
superconductors. \textit{Rev. Mod. Phys.} \textbf{83}, 1057 (2011).

\bibitem{Potter-12prb} Potter A. C. \& Lee, P. A. Topological
superconductivity and Majorana fermions in metallic surface states. \textit{%
Phys. Rev. B} \textbf{85}, 094516 (2012).

\bibitem{sau2} Sau, J. D., Tewari, S. \& Das Sarma, S. Experimental and
materials considerations for the topological superconducting state in
electron- and hole-doped semiconductors: Searching for non-Abelian Majorana
modes in 1D nanowires and 2D heterostructures. \textit{Phys. Rev. B} \textbf{85},
064512 (2012).

\bibitem{xiaodi-mo} Xiao, D., Liu, G.-B., Feng, W., Xu, X. \& Yao, W.
Coupled spin and valley physics in monolayers of MoS$_{2}$ and other
group-VI dichalcogenides. \textit{Phys. Rev. Lett.} \textbf{108}, 196802
(2012).

\bibitem{Mak-10prl} Mak, K. F., Lee, C., Hone, J., Shan, J. \& Heinz, T. F.
Atomically thin MoS$_{2}$: A new direct-gap semiconductor. \textit{Phys.
Rev. Lett.} \textbf{105}, 136805 (2010).


\bibitem{AKis} Radisavljevic, B., Radenovic, A., Brivio, J., Giacometti, V.
\& Kis, A. Single-layer MoS$_{2}$ transistors. \textit{Nature Nanotech.}
\textbf{6}, 147--150 (2011).

\bibitem{cao-nat-com} Cao, T., Feng, J., Shi, J., Niu, Q. \& Wang, E.
Valley-selective circular dichroism of monolayer molybdenum disulphide.
\textit{Nature Commun.} \textbf{3}, 887 (2012).

\bibitem{Mak2} Mak, K. F., He, K., Shan, J. \& Heinz, T. F. Control of
valley polarization in monolayer MoS$_{2}$ by optical helicity. \textit{%
Nature Nanotech.} \textbf{7}, 494--498 (2012).

\bibitem{xiaodongcui} Zeng, H., Dai, J., Yao, W., Xiao, D. \& Cui, X. Valley
polarization in MoS$_{2}$ monolayers by optical pumping. \textit{Nature
Nanotech.} \textbf{7}, 490--493 (2012).

\bibitem{xu1} Wu, S. \textit{et al.} Electrical Tuning of Valley Magnetic
Moment via Symmetry Control in Bilayer MoS$_{2}$. \textit{Nature Phys.}
\textbf{9}, 149-153 (2013).

\bibitem{xu2} Ross, J. S. \textit{et al.} Electrical control of neutral and
charged excitons in a monolayer semiconductor. \textit{Nature Commun.}
\textbf{4}, 1474 (2013).

\bibitem{Helveg} Helveg, S. \textit{et al.} Atomic-Scale Structure of
Single-Layer MoS$_{2}$ Nanoclusters. \textit{Phys. Rev. Lett.} \textbf{84},
951 (2000).

\bibitem{Bollinger} M. V. Bollinger \textit{et al.} One-Dimensional Metallic
Edge States in MoS$_{2}$. \textit{Phys. Rev. Lett.} \textbf{87}, 196803
(2001).

\bibitem{Liuguibin} Liu, G., Shan, W., Yao, Y., Yao, W., \& Xiao, D. A
three-band tight-binding model for monolayers of group-VIB transition metal
dichalcogenides. arXiv:1305.6089.

\bibitem{zhuzhiyong} Zhu, Z. Y., Cheng, Y. C. \& Schwingenschlogl, U. Giant
spin-orbit-induced spin splitting in two-dimensional transition-metal
dichalcogenide semiconductors. \textit{Phys. Rev. B} \textbf{84}, 153402
(2011).

\bibitem{Wang-jacs} Wang, Z. \textit{et al.} Mixed Low-Dimensional
Nanomaterial: 2D Ultranarrow MoS$_{2}$ Inorganic Nanoribbons Encapsulated in
Quasi-1D Carbon Nanotubes. \textit{J. Am. Chem. Soc.} \textbf{132}, 13840
(2010).

\bibitem{Chen-jacs} Li, Y., Zhou, Z., Zhang, S., \& Chen, Z. MoS$_{2}$
Nanoribbons: High Stability and Unusual Electronic and Magnetic Properties.
\textit{J. Am. Chem. Soc.} \textbf{130}, 16739 (2008).

\bibitem{PanHui} Pan, H., and Zhang, Y. Edge-dependent structural,
electronic and magnetic properties of MoS2 nanoribbons. \textit{J. Mater.
Chem.} \textbf{22}, 7280 (2012).

\bibitem{Seifert} Erdogan, E., Popov, I.H., Enyashin, A.N., \& Seifert, G.
Transport properties of MoS$_{2}$ nanoribbons: edge priority. \textit{Eur.
Phys. J. B} \textbf{85}, 33 (2012).

\bibitem{Wang-JMC} Wang, Z. \textit{et al.} Ultra-narrow WS$_{2}$
nanoribbons encapsulated in carbon nanotubes. \textit{J. Mater. Chem.}
\textbf{21}, 171180 (2011).

\bibitem{ZhengLiu} Liu, Z. \textit{et al.} Identification of active atomic
defects in a monolayered tungsten disulphide nanoribbon. \textit{Nature
Commun.} \textbf{2}, 213 (2011).

\bibitem{Tudor-13prb} Stanescu, T., Lutchyn, R. M., \& Das Sarma, S.
Dimensional crossover in spin-orbit-coupled semiconductor nanowires with
induced superconducting pairing. \textit{Phys. Rev. B} \textbf{87}, 094518
(2013).

\bibitem{Xu-acsnano} Wu, S. \textit{et al.} Vapor-Solid Growth of High
Quality MoS2 Monolayers With Near-Unity Valley Polarization. \textit{ACS
Nano.} \textbf{7}, 2768 (2013).

\bibitem{Zande-natm} van der Zande, A. M. \textit{et al.} Grains and grain
boundaries in highly crystalline monolayer molybdenum disulphide. \textit{%
Nat. Mat.} \textbf{12}, 554(2013).

\bibitem{ptype1} Zeng, Z. \textit{et al.} Single-Layer Semiconducting
Nanosheets: High-Yield Preparation and Device Fabrication. \textit{Angew.
Chem. Int. Ed.} \textbf{50}, 11093 (2011).

\bibitem{ptype2} Zhan, Y., Liu, Z., Najmaei, S., Ajayan, P. M., \& Lou, J.
Large-Area Vapor-Phase Growth and Characterization of MoS$_{2}$ Atomic
Layers on a SiO$_{2}$ Substrate. \textit{Small} \textbf{8}, 966 (2012).

\bibitem{Dolui} Dolui, K., Rungger, I. \& Sanvito, S. Origin of the n-type
and p-type conductivity of MoS$_{2}$ monolayers on a SiO$_{2}$ substrate.
arxiv:1301.2491

\bibitem{YZhang} Zhang, Y. \textit{et al.} Ambipolar MoS$_{2}$ Thin Flake
Transistors. \textit{Nano Lett.} \textbf{12}, 1136 (2012).

\bibitem{ELK} http://elk.sourceforge.net/\newline

\end{thebibliography}
\end{document}